\documentstyle[pra,aps,twocolumn]{revtex}

\begin{document}

\title{One and two dimensional Bose-Einstein condesation of atoms
in dark magneto-optical lattices}

\author{A.V. Taichenachev$^{1,2}$, A.M. Tumaikin$^1$ and V.I. Yudin$^{1,2}$}

\address{(1)Laser Physics Laboratory, Novosibirsk State University,\\
Pirogova 2, Novosibirsk 630090, Russia}
\address{(2) Time and Frequency Division,\\
National Institute of Standards and Technology, Boulder CO 80303}

\date{\today}
\maketitle

\begin{abstract}

The motion of atoms in a dark magneto-optical
lattice is considered.  This lattice is formed by a non-uniformly
polarized laser field in the presence of a static magnetic field.
Cold atoms are localized in the vicinity of points where dark states
are not destroyed by a magnetic field. As a result the optical
interaction tends to zero (dark or gray lattices). Depending on the
field configuration such a lattice can exhibit one or
two-dimensional periodicity.  It is shown that in $1D$ and $2D$ dark
magneto-optical lattices the effects connected with the
Bose-statistics of particles can be observed under the
temperature $10^{-5}-10^{-6}K$ and densities $10^{11}-10^{12}
cm^{-3}$, those are usual for current experiments on laser cooling.

\end{abstract}
\vspace{0.5 cm}

\pacs{PACS numbers: 03.75.Fi, 32.80.Pj}


\section{Introduction}
After the outstanding experiments on
Bose-Einstein condensation (BEC) of an atomic gas in a magnetic trap
\cite{anderson,bradley,davis} studies of phenomenon originated
from quantum statistics of particles became one of the main subjects
in contemporary atomic physics. Achieving of BEC by optical means is
of especial interest \cite{chu,co2,solomon,boiron}. On the one hand this
allows deeper understanding of the nature and physical properties of
condensates under various conditions. On the other hand, the works in
this direction may lead to new effective methods of super-deep
cooling. It is worth noting, that in the
refs.\cite{anderson,bradley,davis} laser fields play an auxiliary
role and they are used for precooling down to sufficiently low
temperatures with successive evaporating cooling in a magnetic trap,
when in the final stage of experiments optical fields are absent and
the main part of atoms escape from a trap under evaporative cooling.
One of the main components of the optical methods \cite{chu,solomon}
is the formation of a non-dissipative optical or magneto-optical
potential, when atoms  scatter photons with very low rates.  There
exist two principal ways to solving this problem.  The first one
consists in the use of far-off-resonance light fields with the high
intensity \cite{chu,solomon,jessen}. In this case the potential depth
of order of $10^3\; \varepsilon_r$ ($\varepsilon_r = (\hbar k)^2/2M$
is the single-photon recoil energy) and the rate of spontaneous
photon scattering about $1\; s^{-1}$ are achived
\cite{solomon,boiron}.  Another way is connected with the use of
coherent population trapping (CPT) phenomena in near-resonance light
fields.  As is known \cite{smirnov}, under the resonance interaction
of a polarized radiation with atoms, having optical transitions
$F_g=F \rightarrow F_e=F$ with $F$ an integer and $F_g=F \rightarrow
F_e=F-1$ ($g$ and $e$ denote the ground and excited states
respectively), there exist dark states.  These states are coherent
superposition of the ground-level Zeeman substates, which is fully
decoupled with light $({\bf \widehat{d}}\cdot {\bf E})|\psi_{nc}
\rangle= 0$.  Due to optical pumping atoms are trapped in these
states and do not scatter light.  The use of fields with a
polarization gradient allows to create a potential in dark states
(dark potential). Although for dark states ac Stark shift vanishes,
dark potentials can be created by the atomic translational motion
(so-called gauge or geometric potentials)
\cite{dynamics,dum,nienhuis}, or by applying of a static (magnetic or
electric) field.  In the later case the atomic multipole moments are
spatially non-uniform, that leads to the coordinate dependence of the
interaction energy with a static field. As a result, a periodic
potential is formed. From the other hand, static fields induce
precession of multipole moments and destroy dark states.  However,
this effect can be suppressed to negligible values due up to two
factors:
\begin{enumerate}
\item We can choose a specific geometry
of fields \cite{yudin1} where atoms are localized near the points,
where dark states are not destroyed by a static field.
\item The use of high-intensity laser field allows to lack atoms in
dark states. In the result, as is shown in ref.\cite{qma}, the rate
of spontaneous scattering is inverse proportional to the light
intensity.
\end{enumerate}
A quantitative treatment of dark
magneto-optical lattices in the non-dissipative regime at
high-intensity laser field had been developed in ref.\cite{qma} in
the one-dimensional case. It had been shown that the potential depth
is determined by the ground-state Zeeman splitting $\hbar\Omega$,
while the periodof lattice is of order of the light wavelength
$\lambda$. Thus, we can obtain a very deep potential with a large
spatial gradient. Both these reasons lead to the large energy
separation between vibrational levels $\sim \sqrt{\hbar \Omega
\varepsilon_r}$, which can exceed the laser cooling
temperature.  Under these conditions, atoms being in the lower
vibrational levels can be localized within a very small distance
$\sim \lambda \sqrt{\varepsilon_r/\hbar\Omega}$. We stress that
tunneling between wells is exponentially small (by factor
$\exp(-\sqrt{\hbar \Omega/\varepsilon_r})$), that allows to consider
atoms in each of wells as independent systems.

In the present paper, with $F_g=1
\rightarrow F_e=1$ transition as an example, one- and two-dimensional
non-dissipative magneto-optical lattices are considered with especial
attention to the formation of
atomic structures with lower dimensions ($2D$
-- planes, $1D$ -- lines). In such a lattice the spontaneous
scattering of photons is strongly reduced and the main dissipative
mechanism is elastic interatomic collisions. In the framework of an
ideal Bose-gas model it is shown that under applying of an
additional weak confining potential, the condensation is possible
at the temperatures and densities typical for the current
experiments on laser cooling.

It is worth noting, in the low-saturation limit, that corresponds to
the dissipative regime, dark magneto-optical lattices had been
theoretically \cite{grynberg} and experimentally \cite{hemmerich}
studied.  Sub-Doppler cooling down to $20\; \varepsilon_r$ had been
predicted and observed.  Experimental evidence of the translational
motion quantization had been presented.

\section{Non-dissipative dark magneto-optical lattice}
Let us consider a gas of Bose-atoms with total angular momenta
$F_g=1\rightarrow F_e=1$ in a resonant spatially
nonuniform monochromatic laser field
\begin{equation} \label{field}
{\bf E}({\bf r},t)={\bf E}({\bf r})\: e^{-i\omega t}+c.c.
\end{equation}
in the presence of a static magnetic field ${\bf B}$.
As is known [6], for all $F\rightarrow F$ ($F$ is an integer)
transitions there exist dark CPT-states uncoupled with a laser
field (\ref{field}):
\begin{equation}
\left( \widehat{\bf d}\cdot {\bf E}({\bf r})\right)\, |\psi_{nc}({\bf r})
\rangle =0\: ,
\end{equation}
where $\widehat{\bf d}$ is the dipole moment operator. In the case
under consideration $F=1$ this state has the form \cite{taich}:
\begin{equation} \label{cptstate}
|\psi_{nc}({\bf
r})\rangle=\frac{1}{|{\bf E}({\bf r})|}\, \sum_{q=0,\pm 1}E^q({\bf
r})\, |g,\mu=q\rangle\: ,
\end{equation}
where  $E^q({\bf r})$ are the field components in the
spherical basis
$\{ {\bf e}_0={\bf e}_z,\, {\bf e}_{\pm 1} =\mp ({\bf e}_x\pm
i{\bf e}_y)/\sqrt{2}\}$.
The state (\ref{cptstate}) is a superposition of the
ground-state Zeeman wave functions $|g,\mu\rangle$.
We note that in the general case this state
is neither eigenvector of the Hamiltonian of the interaction
with the magnetic field
$\widehat{H}_B=-(\widehat{\bf \mu}\cdot{\bf B})$
nor of the kinetic energy Hamiltonian
$\widehat{H}_K=\widehat{p}^2/2M$.  Thus, the state $|\psi_{nc}({\bf
r})\rangle$ is not strictly stationary. However, the corrections
to the wave function
resulting from the translational motion and the magnetic field can be
considered as small perturbations with respect to the atom-light
interaction under the conditions:
\begin{equation}
\label{stronglight} V({\bf r})\sqrt{G}\, \gg \,  k\bar{v}, \Omega\, ,
\end{equation}
where $V({\bf r})=|<\hat{d}>{\bf E}({\bf r})|/\hbar$
is the Rabi frequency, $G=V^2({\bf r})/[\gamma^2/4+\delta^2+V^2({\bf
r})]$ is the effective saturation parameter, $\delta$ is the
detuning, $\gamma$ is the inverse lifetime of the excited state
and $\bar{v}$ is the average atomic velocity. In this case the main
part of atoms is pumped into the dark state $|\psi_{nc} ({\bf
r})\rangle$. Under the conditions (\ref{stronglight}) the
relative populations in the CPT-state $n_{nc}$ and in the excited
state $n_e$ obey the relation:
$$
(1-n_{nc}) \sim n_e\sim \left(
\frac{\max\{k\bar{v}, \Omega\} }{V({\bf r}) \sqrt{G}}\right) ^2 \ll
1\, .
$$
Then the evolution of a single atom can be described, with the
same accuracy,  by the effective Hamiltonian:
\begin{equation} \label{onepart}
\widehat{H}^{(1)}_{eff}=\langle\psi_{nc}({\bf
r})|(\widehat{H}_K+ \widehat{H}_B)|\psi_{nc}({\bf r})\rangle\, .
\end{equation}

Using the explicit form of the CPT-state (\ref{cptstate}), we write
the one-particle Hamiltonian (\ref{onepart}) as a sum of four terms:
\begin{equation} \label{eff}
\widehat{H}^{(1)}_{eff}=\frac{\widehat{p}^2}{2M}+
U({\bf r})+\frac{1}{2M}\left\{({\bf A}({\bf r})\cdot\widehat{\bf p})+
(\widehat{\bf p}\cdot{\bf A}({\bf r}))\right\}+W({\bf r})\,.
\end{equation}
The first term is the kinetic Hamiltonian. The second one is the
magneto-optical potential:
\begin{equation} \label{mopot}
U({\bf r})=\hbar\Omega\,\frac{i({\bf B}\cdot[{\bf E}({\bf r})\times
{\bf E}^{\ast}({\bf r})])}{|{\bf B}|\, |{\bf E}({\bf r})|^2}\, ,
\end{equation}
which is independent of the amplitude and phase of light field.
The last two corrections in Eq.(\ref{eff}) caused by
the translational motion of atom. The first of these
is of the order of $kv$ and can be interpreted as the interaction
with the effective vector-potential:
\begin{equation}  \label{vecpot}
A_j({\bf r})=-i\hbar \left(\frac{{\bf E}^{\ast}}{|{\bf
E}|}\cdot\frac{\partial}{\partial x_j}\frac {{\bf E}}{|{\bf
E}|}\right)\,.
\end{equation}
The second correction is of the order of the recoil energy
$\hbar\omega_r$ and makes a contribution into the atomic potential
energy:
\begin{equation} \label{recpot}
W({\bf r})=\frac{\hbar^2}{2M}\sum_{j}
\left|\frac{\partial}{\partial x_j}\frac{{\bf E}}{|{\bf E}|}\right|^2\,.
\end{equation}
If the Zeeman splitting obeys the conditions
\begin{equation} \label{zeeman}
\Omega\,\gg\, k\bar{v}, \varepsilon_r/\hbar \;\; ,
\end{equation}
then the last two terms in Eq.(\ref{eff}) are negligible, i.e.
$$
\widehat{H}^{(1)}_{eff}\approx\frac{\widehat{p}^2}{2M}+U({\bf r})\, .
$$
In this case the problem is reduced to the motion of a particle
in the magneto-optical potential (\ref{mopot}) only. The depth of
this potential is determined by
the ground-state Zeeman
splitting $\hbar \Omega$ (below we suppose  $\Omega > 0$)
and its
period is of order of the light wavelength $\lambda$.

As is well known, in a periodic potential the energy spectra has the
band structure. However,  due to the condition
(\ref{zeeman}) the tunneling is negligible for the lower bands,
i.e. the strong binding
of the particle in a single well is realized. It can be shown that the
widths of the lower bands are exponentially small by the factor
$\exp\left(-\sqrt{\hbar\Omega/\varepsilon_r}\,\right)$
with respect to the energy separation between bands.
The positions of these bands are determined (with good accuracy)
by a harmonic expansion of the potential in the vicinity of the well
bottom:
\begin{equation} \label{mopot2}
U({\bf r})\approx \hbar\Omega k^2
\sum_{i,j=1,2,3} C_{ij}x_i x_j \,.
\end{equation}
As is seen, the separation between lower levels is of the order of
$\sqrt{\hbar\Omega\varepsilon_r}$  at the eigenvalues
of $\widehat{C}$ of order of $1$.

As it has been shown in ref.\cite{qma}, atoms being in the lower
vibrational levels scatter photons with extremely low rates:
$$
\tau^{-1} \sim \gamma \left(\frac{\Omega}{V}\right)^2
\sqrt{\frac{\varepsilon_r}{\hbar \Omega}}\ll\gamma \;\;.
$$
Here the factor $(\Omega/V)^2 \ll 1$ is directly connected with
CPT-effect, when in a strong light field the probability of leaving
of a dark state is inverse proportional to the light intensity. The
additional multiplier $\sqrt{\varepsilon_r/\hbar\Omega}\ll 1$ arises
from the localization of atoms in the vicinity of points, where dark
state are not destroyed by a magnetic field.

\section{Ideal Bose-gas in dark magneto-optical lattices}
As it has been indicated in the Introduction, the spontaneous
photon scattering in dark magneto-optical lattices can be strongly suppressed.
Hence the main dissipative mechanism, leading to the thermal
equilibrium, is the elastic interatomic collisions.  This allows to
apply the methods of statistical physics to study of a stationary
state of atoms, which corresponds to the thermodynamic equilibrium.
In the present paper we restrict our treatment by the ideal gas
model when the contribution of atom-atom interactions into the
system energy is negligible.  At the same time collisions are
implicitly taken into account as a reason of the thermal equilibrium.

BEC is one of the most interesting phenomenon arising in a Bose
 gas under sufficiently low temperatures.  As it has been shown
 by recent studies, the character and parameters of the phase
 transition essentially depend on the system dimensions and on
 the presence of an external confining potential.  So, it is
 well-known \cite{huang}, that in the case of a free gas BEC in
 the 1D and 2D cases is absent.  However, as it has been
recently shown in refs.\cite{kleppner,ketterle} in both 1D and 2D
cases BEC becomes possible under applying a confining potential.
 Moreover, the conditions for the BEC achievement are appreciably
 less stringent than in the 3D case.

 The non-dissipative dark magneto-optical lattice considered in the previous
section are promising tools for studies of BEC in systems with
lower dimensions.  Let us consider the concrete examples.

\subsection{1D lattice -- 2D condensation}
The simplest
realization of 1D dark magneto-optical lattice is the $lin \perp lin$ light
field configuration plus a magnetic field directed along the wave
prorogation direction (see in fig.2).  Here the dark magneto-optical potential
has the form \cite{qma}:
 \begin{equation} \label{lplpotential}
 U=- \hbar \Omega \cos (2 kz) \; .
 \end{equation}
Atoms are localized in the planes $z_n = \lambda n/2$, where the
field has the $\sigma_{-}$ polarization
 and the dark state
coincides with the Zeeman substate \mbox{$|F_g=1, \mu=-1 \rangle$}.  The
lower energy levels of the potential (\ref{lplpotential})
correspond to the localization of atoms in the vicinity of the single
 well bottom, when tunneling between wells is negligible.  Basing on
a harmonic approximation, one can find the energy separation between
the lower levels
$\Delta \varepsilon = \sqrt{8\hbar\Omega \varepsilon_r}$.
Then under the temperatures $$ k_B T < \sqrt{8\hbar
\Omega \varepsilon_r} $$ atoms are in the vibrational ground
state.  In the other words, the translational motion of atoms along
$z$ is frozen.  For instance, for the D1-line of $^{87} Rb$ ($\lambda
= 787\, nm$) under the magnetic field amplitude $B \sim 4\, G$
freezing is achived at $T < 10^{-5} K$.  Due to the absence of
tunneling, each of localization planes can be considered as an
independent thermodynamic and mechanical system, where particles
freely move along the $x$ and $y$ axes.

If an additional confining (in the $x y$-plane) potential is
applied (for example, far-off-resonance optical shift), then BEC
can be reached in a single plane.  As it was shown in
ref.\cite{ketterle}, for 2D harmonic
potential with the frequency $f$ the expression for the
transition temperature is given by:
\begin{equation} \label{2Dpoint}
N = 1.6 \left(\frac{k_B T_c}{\hbar f}\right)^2 \;,
\end{equation}
where $N$ is the number of atoms in a single plane.  For an
atomic sample with the density $n \sim 10^{11}-10^{12} cm^{-3}$ and
with the size $L \sim 10^{-1} cm$ we have $N \sim 10^{5}-10^{6}$ for
each plane at the periodicity about $10^{-4} cm$.  Then for a
confining potential with $f \sim 10^{3} Hz$ the transition
temperature $T_c \sim 10^{-5} K$ is in few orders higher than the
transition temperature in 3D magnetic traps
\cite{anderson,bradley,davis}.

In the case under consideration BEC can be observed, for
instance, by the time-of-flight measurements of the atomic
momentum distribution after turning off of a confining potential.
It should be noted that the phases of condensates in each planes are
independent. So, if the lattice is considered as whole, we have the
quasicondensation only.

\subsection{2D lattice -- 1D condensation}
The example of three-beam field configuration where the 2D dark magneto-optical
potential is formed is shown in fig.3. Here the three wave-vectors
lie in the $xy$-plane and make an angle $2 \pi/3$ one with another.
The linear polarizations of beams make the same angle $\phi\neq
0,\,\pi/2$ with the $z$-axis. This angle $\phi$ can be varied in a
wide domain. The main reason for the inequality $\phi \neq \pi/2$ is
the fact that in the case of $\phi=\pi/2$ there exist lines, where
the field amplitude is equal to zero due to the interference. In the
vicinity of these lines the CPT conditions (\ref{stronglight}) are
violated. The case of $\phi=0$ is not suitable due to the absence of
magneto-optical potential (\ref{mopot}).

Atoms are localized in the vicinity of straight lines, where the field
has $\sigma_{-}$ circular polarization. In the same manner as in the
previous section, one can show that under sufficiently low
temperatures $k_B T < \sqrt{\hbar \Omega \varepsilon_r}$ the
translational degrees of freedom along $x$ and $y$ are frozen out.
Each of lovalization lines can be considered as an independent 1D
system.  Under applying of a confining (along $z$) harmonic potential
we have the condensation at the temperature \cite{ketterle}:
\begin{equation}
\label{1Dpoint} N = \frac{k_B T_c}{\hbar f}\log\left(\frac{2 k_B
T_c}{\hbar f}\right)\;,
\end{equation}
where $N$ is the number of atoms in a line, $f$ is the oscillation
frequency.

For an atomic sample with the density $n \sim 10^{12} cm^{-3}$ and
the size $L \sim 10^{-1} cm$ we have $N \sim 250$ for each line at
the periodicity about $10^{-4} cm$. Then under $f \sim 10^{3} Hz$ we
find $T_c \sim 10^{-6} K$, that is typical for laser cooling
experiments.

Note that like the case of 1D lattice, here
the quasicondensation is possible only, if the whole gas volume is
considered.

\section{Conclusion}
We have considered dark magneto-optical lattices in the
regime, when the Rabi frequency of laser field is much greater than
the Zeeman splitting of the ground state. In such a regime the
lattice is essentially non-dissipative. We have shown that $2D$ and
$1D$ atomic structures can be formed in these lattices. Then we have
studied the Bose-Einstein condensation in the lower-dimensional
systems in the framework of the ideal gas model. It was predicted
that BEC is possible in both $2D$ and $1D$ cases under the
temperatures $10^{-6}-10^{-5} K$ and densities $10^{11}-10^{12}
cm^{-3}$.
Concluding we note that the above developed approach
can be applied to any non-dissipative optical lattice with
the sufficient large depth. The example is a far-off-resonance
optical lattice \cite{co2,jessen}.

\acknowledgments

Authors thank to Leo Hollberg and all members of his scientific group
at NIST, Boulder for helpful discussions.
AVT and VIYu acknowledge the hospitality of NIST, Boulder.
This work was supported by
the Russian Foundation for Basic Research (grant no. 98-02-17794).


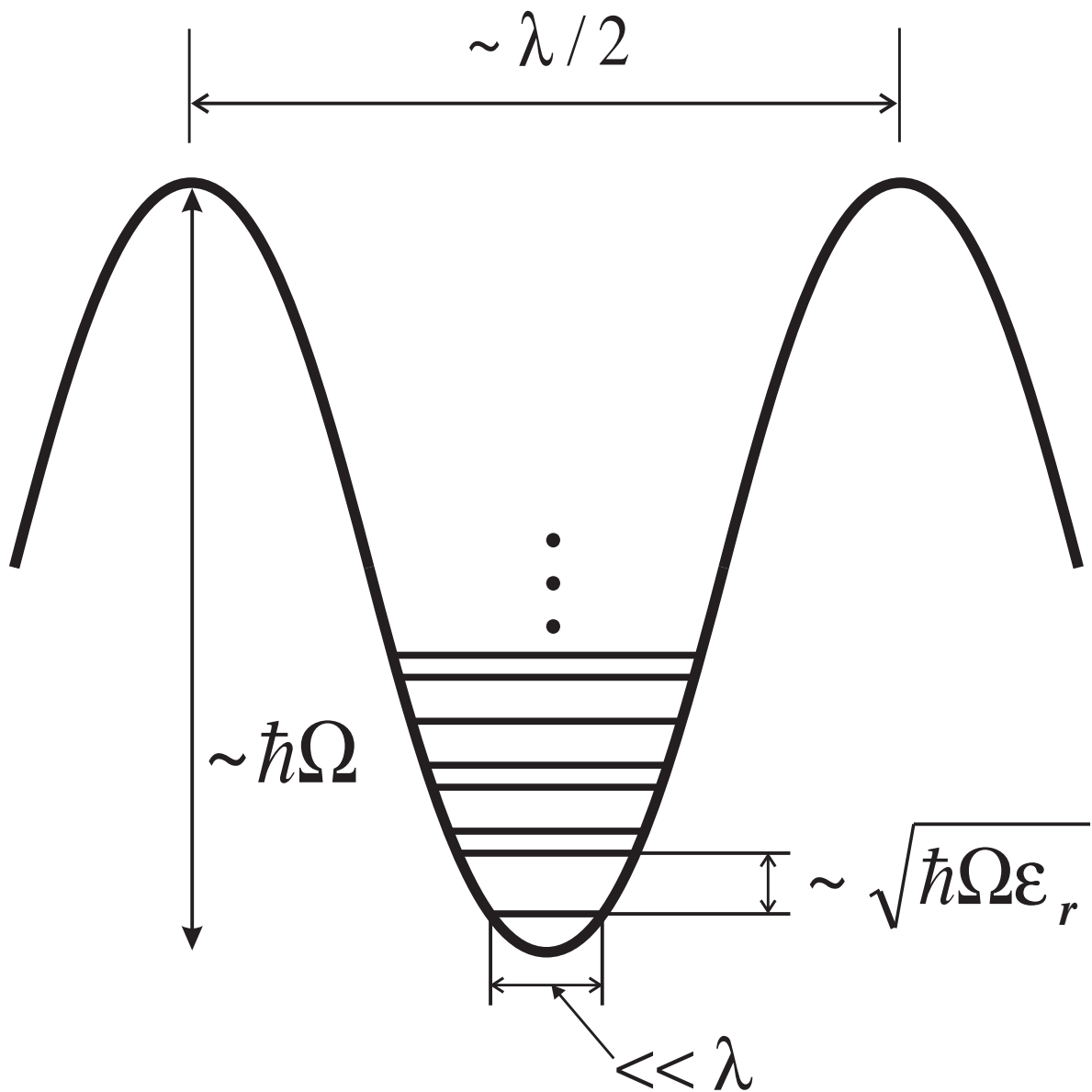
\begin{figure}
\caption{Sheme of lower-energy vibrational strusture in a dark
magneto-optical lattice.}
\end{figure}

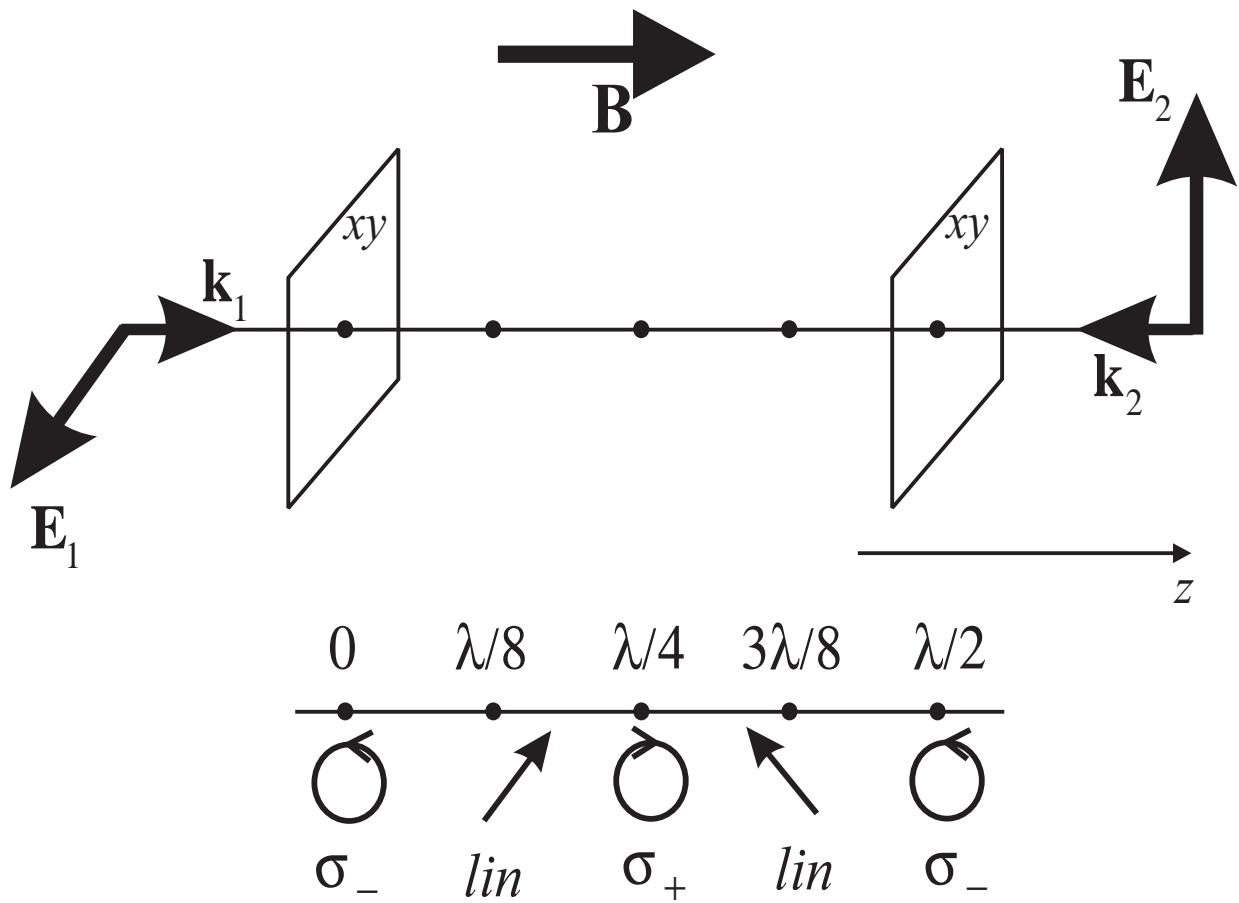
\begin{figure}
\caption{The $lin\perp lin$ light field configuration, when
one-dimensional dark magneto-optical potential is fromed. Atoms
freely move in the planes (parallel to the $x y$-plane), where the
field has the $\sigma_{-}$ polarization.} \end{figure}

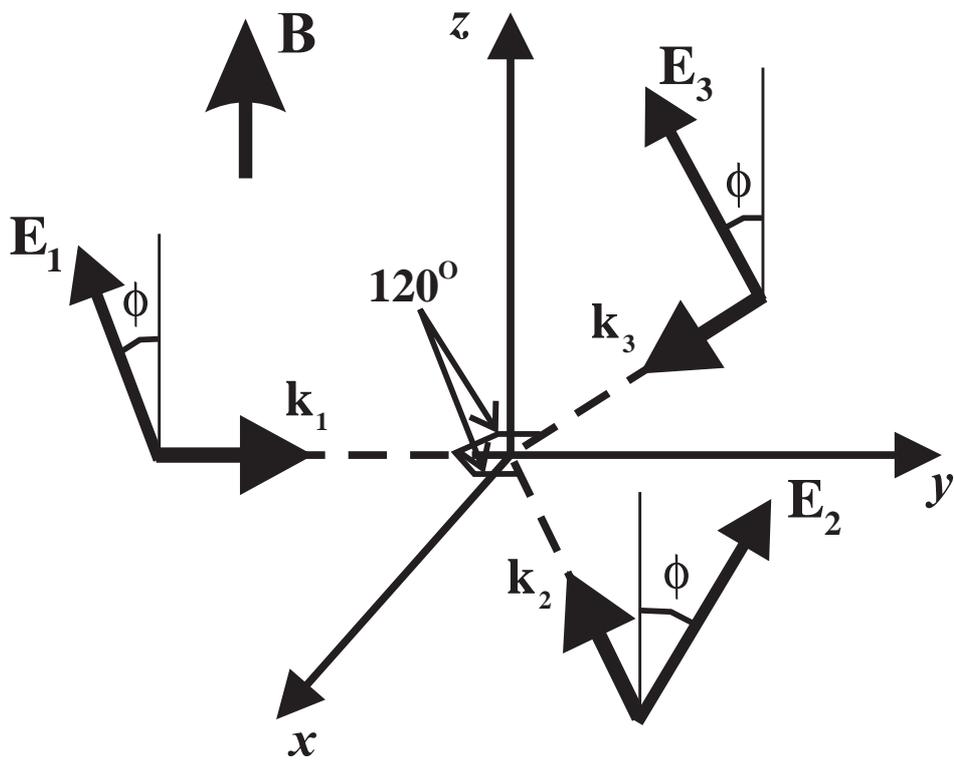
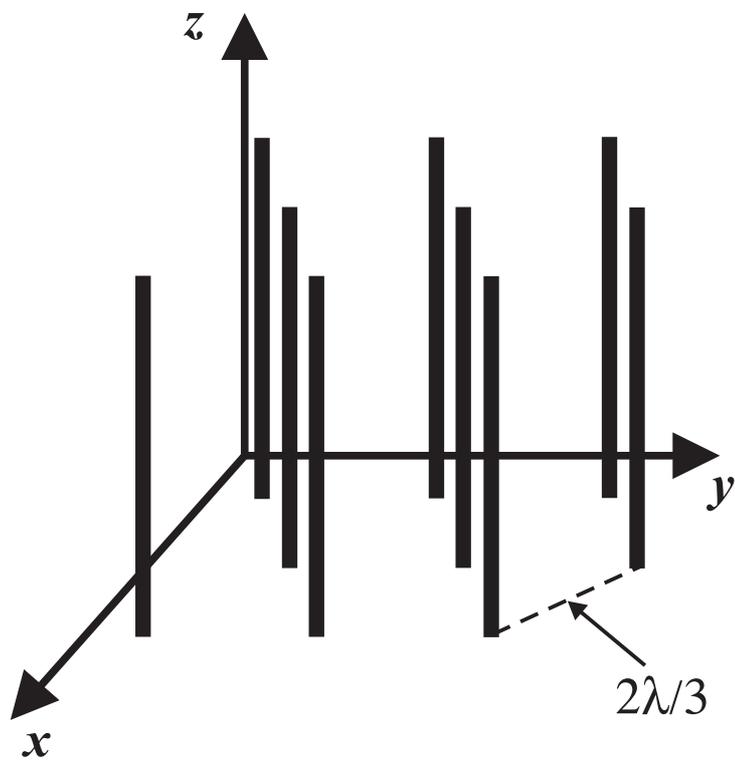
\begin{figure}
\caption{The three-beam light field configuration, when
two-dimensional dark magneto-optical potential is fromed.
Atoms freely move in the lines (parallel to the $z$-axis), where the
field has the $\sigma_{-}$ polarization.}
\end{figure}

\end{document}